\documentclass[openany]{report}
\usepackage{graphicx}
\usepackage{natbib}
\usepackage{xcolor}
\usepackage{hyperref}

\title{{\tt RHODIUM}: A post-processor for  {\tt BIGSTICK} configuration-interaction wave functions} 

\author{Calvin W. Johnson\footnote{Department of Physics, San Diego State University, 5500 Campanile Drive, San Diego CA 92182-1233}}

\begin{document}

\maketitle

\begin{abstract}
{\tt RHODIUM} is a postprocessing code for nuclear structure physics. 
%This is the manual for {\tt RHODIUM}, a postprocessing code to be used with {\tt BIGSTICK}. 
It can be used to compute density matrices, spectroscopic amplitudes, and other information, from wave function and basis files created by the configuration-interaction shell-model code  {\tt BIGSTICK}. The source code is available at {\tt github.com/cwjsdsu/Rhodium}. This manual gives detailed instructions how to use.
\end{abstract}

\tableofcontents

%\part{Introduction}

\chapter{Introduction}

{\tt BIGSTICK} (\cite{johnson2018bigstick,Johnson20132761}) is an efficient and powerful configuration-interaction code for tackling the many-fermion problem, primarily modeling atomic nuclei. In addition to finding the eigenpairs of the many-nucleon system, 
{\tt BIGSTICK} can carry out a variety of additional tasks, such as computing expectation values, transition density matrices, and so on. 

{\tt BIGSTICK}  has one key restriction, which both enables its efficiency but also limits the tasks it can carry out: it uses a basis with certain fixed quantum numbers, such as total $M/J_z$ ($z$-component of angular momentum) and $N$ and $Z$ (number of valence neutrons and protons, respectively).  Thus, {\tt BIGSTICK} cannot compute spectroscopic amplitudes, which are matrix elements of adding or removing one nucleon, nor 
of charge-changing transitions such as $\beta$-decay (or only through some assumptions such as isospin rotation). Note that one has the option to fix parity or to allow for both parities simultaneously.

{\tt RHODIUM} addresses this limitation. {\tt RHODIUM} reads in files generated by {\tt BIGSTICK}, but unlike 
{\tt BIGSTICK}, it can carry out operations between wave functions computed in two different bases with different quantum numbers. In particular, {\tt RHODIUM} can:
\begin{itemize}
\item Compute the one-body density between wave functions (the name {\tt RHODIUM} refers both to the symbol $\rho$ for density matrices, as well as being a \textit{dense} metal), even if the bases have different $M$ and/or different parity $\pi$; charge-changing densities can also be computed;

\item Apply a general one-body operator to a wave function (e.g., E1, E2, or $\hat{J}_+$) or set of wave functions. The final state(s) can be in a basis with different quantum numbers, e.g., $M$ or $\pi$, from the initial state(s).  They can also have different $T_z$.

\item Compute the spectroscopic amplitudes between two wave functions with different particle number;

\item Add or remove a particle or superposition of particles to a wave function;

    \item Compute the overlap between two wave functions with different many-body truncations;

    \item Project a wave function existing in one basis into another basis with a different truncation;

\end{itemize}
and so on.

The resulting wave functions are written back as files in a format usable by {\tt BIGSTICK}.  One can summarize the difference between {\tt BIGSTICK} and {\tt RHODIUM} as {\tt BIGSTICK} is much more efficient, but limited to a fixed basis, while {\tt RHODIUM} is more flexible, able to connect two different bases, but less efficient. {\tt RHODIUM} does not diagonalize a Hamiltonian and currently does not deal with two-body operators.

\section{Current version}

The current version of the code this manual has been updated for is 1.0.0.

\bigskip

\textbf{NOTE}: This manual is still in progress. Some options may not be fully addressed.

Parallelization is in process. At this time, only OpenMP parallelization is implemented, and is not yet very efficient.  Future versions should have improved parallelization.  Although MPI is a compile option, it does not fully work.

\section{License}

This code is distributed under the MIT Open Source License. 

The source code and sample inputs are found at 
 {\tt github.com/cwjsdsu/Rhodium}.

 The {\tt BIGSTICK} code can be found at {\tt github.com/cwjsdsu/BigstickPublick}.

Permission is hereby granted, free of charge, to any person obtaining a copy of this software and associated documentation files (the "Software"), to deal in the Software without restriction, including without limitation the rights to use, copy, modify, merge, publish, distribute, sublicense, and/or sell copies of the Software, and to permit persons to whom the Software is furnished to do so, subject to the following conditions:

\medskip

The above copyright notice and this permission notice shall be included in all copies or substantial portions of the Software.

\medskip

THE SOFTWARE IS PROVIDED "AS IS", WITHOUT WARRANTY OF ANY KIND, EXPRESS OR IMPLIED, INCLUDING BUT NOT LIMITED TO THE WARRANTIES OF MERCHANTABILITY, FITNESS FOR A PARTICULAR PURPOSE AND NONINFRINGEMENT. IN NO EVENT SHALL THE AUTHORS OR COPYRIGHT HOLDERS BE LIABLE FOR ANY CLAIM, DAMAGES OR OTHER LIABILITY, WHETHER IN AN ACTION OF CONTRACT, TORT OR OTHERWISE, ARISING FROM, OUT OF OR IN CONNECTION WITH THE SOFTWARE OR THE USE OR OTHER DEALINGS IN THE SOFTWARE.

\subsection{LAPACK copyright notice}

We use LAPACK subroutines in our code.  The following are the LAPACK copyright notices.

\smallskip

Copyright (c) 1992-2013 The University of Tennessee and The University
                        of Tennessee Research Foundation.  All rights
                        reserved.
\smallskip
                        
Copyright (c) 2000-2013 The University of California Berkeley. All
                        rights reserved.

\smallskip

Copyright (c) 2006-2013 The University of Colorado Denver.  All rights
                        reserved.

\smallskip

Additional copyrights may follow

\bigskip

Redistribution and use in source and binary forms, with or without
modification, are permitted provided that the following conditions are
met:

\smallskip

\noindent - Redistributions of source code must retain the above copyright
  notice, this list of conditions and the following disclaimer.

\smallskip

\noindent - Redistributions in binary form must reproduce the above copyright
  notice, this list of conditions and the following disclaimer listed
  in this license in the documentation and/or other materials
  provided with the distribution.
\smallskip

\noindent 

- Neither the name of the copyright holders nor the names of its
  contributors may be used to endorse or promote products derived from
  this software without specific prior written permission.

\bigskip

The copyright holders provide no reassurances that the source code
provided does not infringe any patent, copyright, or any other
intellectual property rights of third parties.  The copyright holders
disclaim any liability to any recipient for claims brought against
recipient by any third party for infringement of that parties
intellectual property rights.

\smallskip

THIS SOFTWARE IS PROVIDED BY THE COPYRIGHT HOLDERS AND CONTRIBUTORS
"AS IS" AND ANY EXPRESS OR IMPLIED WARRANTIES, INCLUDING, BUT NOT
LIMITED TO, THE IMPLIED WARRANTIES OF MERCHANTABILITY AND FITNESS FOR
A PARTICULAR PURPOSE ARE DISCLAIMED. IN NO EVENT SHALL THE COPYRIGHT
OWNER OR CONTRIBUTORS BE LIABLE FOR ANY DIRECT, INDIRECT, INCIDENTAL,
SPECIAL, EXEMPLARY, OR CONSEQUENTIAL DAMAGES (INCLUDING, BUT NOT
LIMITED TO, PROCUREMENT OF SUBSTITUTE GOODS OR SERVICES; LOSS OF USE,
DATA, OR PROFITS; OR BUSINESS INTERRUPTION) HOWEVER CAUSED AND ON ANY
THEORY OF LIABILITY, WHETHER IN CONTRACT, STRICT LIABILITY, OR TORT
(INCLUDING NEGLIGENCE OR OTHERWISE) ARISING IN ANY WAY OUT OF THE USE
OF THIS SOFTWARE, EVEN IF ADVISED OF THE POSSIBILITY OF SUCH DAMAGE.

\chapter{Compiling the code}

The source code and sample inputs are found at 
 {\tt github.com/cwjsdsu/Rhodium}.

 The {\tt BIGSTICK} code can be found at {\tt github.com/cwjsdsu/BigstickPublick}.

The code can be compiled with either the intel {\tt ifort} or the GNU {\tt gfortran} compiler.
It does not require external libraries or packages; it carries with it all the libraries it needs.

{\tt RHODIUM} has  OpenMP parallelism, but it is not efficient in all cases.  We have partial MPI parallelism, but in its current 
version it is not yet fully implemented.

You can find the compile options by 
\begin{verbatim}
make help
\end{verbatim}
The current options are:
\begin{verbatim}
Here are some compile options:
Note default compiler is intel ifort compiler
----------------------------------------------
default compiler is intel ifort
make serial -- default serial, rhodium.x
make openmp -- OpenMP parallelism, rhodium-openmp.x
make mpi -- MPI parallelism, rhodium-mpi.x
make openmp-mpi -- hybrid OpenMP+MPI, rhodium-mpi-omp.x
-----------------------------------------------
make gfortran -- serial with GNU compiler gfortran, rhodium.x
make gfortran-openmp -- OpenMP with GNU compiler gfortran, rhodium-openmp.x
\end{verbatim}
Currently, the recommended compile option is the last one: {\tt make gfortran-openmp}. In the near future we plan to fully implement MPI distributed memory parallelism.

\chapter{Required input}

Here I discuss the required input. 

To prepare  for a calculation, {\tt RHODIUM} needs files with basis information for both the initial and final state(s). These are files with extension {\tt .bas} and are created by {\tt BIGSTICK} using the initial menu option {\tt `(b)'}. If the initial and final spaces are the same, you may reuse the same {\tt .bas} file.

In all applications you need an initial wave function file with extension {\tt .wfn}; this can be generated by {\tt BIGSTICK}. In some applications, {\tt RHODIUM} will create a new final wave function file, which can then either be read back into {\tt BIGSTICK} or perhaps reused by {\tt RHODIUM}. In many applications, but not all, you will also need a final wave function file. \textbf{The initial and final wave functions files must be different files}; if the information is the same, you must make a copy with a different name.  (This is because both 
the initial and the final 
wave function files are kept open simultaneously.)

\section{Creating a basis file with {\tt BIGSTICK}}

Here is how a {\tt BIGSTICK} session creating a basis looks like:
\begin{verbatim}
  Enter choice 
b
  Create basis (.bas) file for later postprocessing 
  
  Enter file with s.p. orbit information (.sps or .sp)
  (Enter "auto" to autofill s.p. orbit info for no-core shell model ) 
  (Enter "?" for more information ) 
sd
  Enter # of valence protons (max  12 ), neutrons (max  12)
4 4
  Enter 2 x Jz of system 
0
  
  .... Building basis ... 
  
  Information about basis: 
                  495  SDs for species            1
                  495  SDs for species            2
  Total basis =                 28503
  
  .... Basis built ... 
  This will write information on the basis useful for postprocessing 
  (Written as a binary file, much like a .wfn file)
  Enter name of file (without extension )
mg24
  assuming only a single fragment 
  finished writing proton SDs 
  finished writing neutron SDs 
  Finished writing to file     
\end{verbatim}
This creates a binary file {\tt mg24.bas}.
(If you want to have an explicit, human-readable file that includes the basis, you can use the option `{\tt (t)}' in {\tt BIGSTICK} to create a 
{\tt .trwfn} file. That file is not usable in {\tt RHODIUM}; you can, however, use it to write your own postprocessing code.

\section{Creating a wave function file}

There are multiple ways to get to a wavefunction file with extension {\tt .wfn}. The first and simplest way is to carry out a default or normal run in {\tt BIGSTICK} with initial menu option `{\tt (n)}'.  Any file with extension {\tt .wfn}, whether created by {\tt BIGSTICK} or {\tt RHODIUM}, should be usable by {\tt RHODIUM} (or can be fed back into {\tt BIGSTICK}). The main difference is that {\tt RHODIUM} requires {\tt .bas} basis files, while {\tt BIGSTICK} `knows' how to construct the basis.

\section{Other inputs}

Other inputs are created by the user as needed, described in subsequent chapters.  These are files that include matrix elements of operators applied to a wave function, such as a {\tt .opme} file for one-body operators, or a {\tt .spa} file for single-particle addition/removal.

\chapter{Density matrices}

Much of what {\tt RHODIUM} does 
is computing `density matrices,' 
which I now discuss in some depth. 
This chapter only discusses the formal 
definition(s) of density matrices. Later 
chapters will guide you through computing them
using {\tt RHODIUM}.

There are many objects called `density matrices' in quantum systems. 
Here we deal with one-body density matrices, which generically have the form
\begin{equation}
   \rho_{ab} \sim \langle \Psi_f | \hat{c}^\dagger_a \hat{c}_b | \Psi_i \rangle,
\end{equation}
where $\hat{c}^\dagger_a, \hat{c}_b$ are nucleon creation and annihilation operators, respectively, creating and destroying nucleons in orbitals $a$ and $b$. 
If $ | \Psi_f \rangle \neq | \Psi_i \rangle$, this is a transition density matrix. Transition density matrix elements can be used, for example, for 
computing $\beta$-decay and $\gamma$-decay probabilities.  On the other hand, if
$ | \Psi_f \rangle = | \Psi_i \rangle$, 
then one can extract either static properties, such as magnetic dipole (M1) or electric quadrupole (E2) moments, or elastic scattering, say of electrons, neutrinos, or dark matter.

If the initial and final numbers of protons and neutrons are the same, then one can compute (charge-conserving) density matrices. 

For one-body operators, one generally 
defines the density matrix as
\begin{equation}
    \rho_K^{f,i}(a,b) = \frac{
    \langle \Psi_f: J_f||  [\hat{c}^\dagger_a \otimes \tilde{c}_b ]_K || \Psi_i:J_i \rangle
    } {\sqrt{2K+1}}. \label{rhodefn}
\end{equation}
This means the initial state $\Psi_i$ has angular momentum $J_i$, the final state $\Psi_f$ has angular momentum $J_f$, and the nucleon creation/annihilation operators are coupled up (through a Clebsch-Gordan coefficient) to total angular mometum $K$. The indices $a,b$ label single-particle orbitals, such as $0d_{5/2}$. The notation $\tilde{c}$ denotes a 
``time-reversed'' operator:
\begin{equation}
    \tilde{c}_{j,m}= (-1)^{j+m}\hat{c}_{j,-m}
\end{equation}
This is necessary to make the operators transform as ``spherical tensors'' (see \citet{edmonds1996angular} for details) as well as to correctly preserve selection rules.
Finally the reduced matrix element (see Edmonds) is defined by 
\begin{eqnarray}
    \langle J_f || \hat{O}_K || J_i \rangle  & = &  (-1)^{J_f -M_f} \frac{\langle J_f M_f | \hat{O}_{K,M_K}
    | J_i M_i \rangle}
    {
\left ( \begin{array}{ccc}
J_f & K & J_i \\
-M_f & M_K & M_i
\end{array}
    \right )
    } \\
    & = & (-1)^{2K}\sqrt{2J_f +1}
    \frac{\langle J_f M_f | \hat{O}_{K,M_K}
    | J_i M_i \rangle}
    {
\left ( J_i M_i, K M_K | J_f M_f 
    \right )
    }. \label{reduced2}
\end{eqnarray}
Note that the phase $(-1)^{2K}$ in Eq.~(\ref{reduced2}) only 
has an effect if $K$ is half-integer, that is, the operator $\hat{O}$ has an odd number of fermion operators, for example to 
create or destroy a particle. 
The reduced matrix element arises from the Wigner-Eckart theorem, which states that any dependence upon orientation, that is, initial $M_i$ and final $M_f$,
must be proportional to a Clebsch-Gordan coefficient.
With the above 
definition, a useful symmetry relation is 
\begin{equation}
\rho_K^{if} (b,a) = (-1)^{j_a - j_b + J_i -J_f} \rho_K^{fi}(a,b).
\end{equation}
    
The reason for the definition Eq.~(\ref{rhodefn}) is to make easy
computing the reduced matrix element of 
an arbitrary one-body operator:
\begin{equation}
\left \langle \Psi_f \left | \left | 
\hat{\cal O}_K \right | \right | \Psi_i \right \rangle 
= \sum_{ab} \rho^{(fi)}_K(a,b)  \langle a || \hat{\cal O} || b \rangle. \label{transitioncalc}
\end{equation}
A number of post-processing codes use one-body densities, for example to compute electromagnetic transitions.

Though {\tt BIGSTICK} computes density matrices very efficiently, it has limitations.  For example, 
{\tt BIGSTICK} works in the $M$-scheme, that is, a fixed $J_z$ or $M$ basis. Therefore, to compute density matrices, {\tt BIGSTICK} actually computes
$\langle J_f M_f | \hat{O}_{K,M_K} | J_i M_i \rangle$ 
and then divides by the appropriate Clebsch-Gordan coefficient to obtain the reduced density matrix element. Some Clebsch-Gordan coefficients vanish, however, usually due to some symmetry.
In that case, the associated reduced density matrix element cannot be computed. This mostly happens for $M=0$. Specifically,
\begin{equation}
   ( J_i M_i, K M | J_f M_f )= 
   (-1)^{J_i + K_i - J_f}
   ( J_i\, -M_i, K\, -M | J_f\, -M_f )
\end{equation}
When $M_i = M_f = M = 0$, then implies that 
the Clebsch-Gordan coefficient must vanish when 
$J_i + K - J_f$ is odd. 

One inelegant solution is to simply re-run {\tt BIGSTICK} with $M \neq 0$. For large cases this can be time-consuming. Furthermore, there can be arbitrary signs introduced.  {\tt RHODIUM} fixes this problem by applying a raising operator $\hat{J}_+$ to change an $M=0$ state to a $M=1$ state.
This breaks the symmetry and allows one to extract the matrix elements missing due to vanishing Clebsch-Gordan coefficients. 

\textbf{It's important to remember that we must have $K \geq |M| = |M_i-M_f|$.} {\tt RHODIUM} will automatically skip such density matrix elements.

Another limitation of {\tt BIGSICK}, and addressed in {\tt RHODIUM}, is the computation of density matrix elements between levels of opposite parity.  {\tt BIGSTICK} allows one to choose positive ($+$) or negative ($-$) parity, or both, ({\tt `0'}). Computing both parities simultaneously allows one to extract cross-parity matrix elements, but at the price of a larger (roughly $2\times$) basis dimension.  {\tt RHODIUM} allows one to compute density matrix elements directly between states of different parities.

\chapter{Running {\tt RHODIUM}}

No matter what action you are taking, the first step is to read in the basis information files, with 
extension {\tt .bas}, created by {\tt BIGSTICK} with initial menu option {`{\tt (b)}'}. Basis files are need for both the initial and final wave functions, but these can be the same. {\tt RHODIUM}  provides some confirming information, such as the basis dimension, valence number, and so on. (Note: as of version 0.9.1, {\tt RHODIUM} can now utilize the hole formalism available in {\tt BIGSTICK}, 
where entering, e.g., `-4' means four holes in a space. One must, however, 
use it consistently, that is, if one space uses the hole formalism, the other space must as well.)
\begin{verbatim}
  Enter name of INITIAL .bas file 
mg24
  Basis file successfully opened 
  about to read in basis
 dimbasischeck=                28503
  Valence Z, N =            4           4
  Single particle space : 
    N    L  2xJ 
    0    2    3
    0    2    5
    1    0    1
....
  DONE READING IN 
  Enter name of FINAL .bas file 
mg24
  Basis file successfully opened 
 dimbasischeck=                28503
  Valence Z, N =            4           4
  Single particle space : 
...
\end{verbatim}
Depending upon the particle numbers, {\tt RHODIUM}
will automatically provide appropriate options. 

For example, if both initial and final spaces have the same number of protons and neutrons, here are the 
options provided:
\begin{verbatim}
  Choose one of the following menu options 
   
  (dotwfn) Dot product between wave functions in different bases (different truncations)
  (dn1iso) 1-body density matrices, non-charge changing, good isospin 
  (dn1xpn) 1-body density matrices, non-charge changing, explicit proton-neutron 
  (ent1bd) 1-body entropy 
  (app1bd) Apply non-charge-changing 1-body operator to wave functions and write out
  (projct) Project a state into a basis with different truncation 
  (lincom) construct linear combinations of wave functions 
   
  Enter the six-character code for your choice::
\end{verbatim}
If the total number of nucleons is conserved, but 
$N$ and $Z$ both change by 1, {\tt RHODIUM} provides 
charge-changing options:
\begin{verbatim}
 (dn1bcc) Charge-changing 1-body densities (pn)
  (d1iscc) Charge-changing 1-body densities (iso)
  (app1cc) Apply charge-changing 1-body operator to wave functions and write out     
\end{verbatim}
Finally, if there is a difference by one in 
$Z$ or $N$, but not both, the options are
\begin{verbatim}
  (spamp1) 1-particle spectroscopic amplitude (change in number of particles)
  (appsp1) Add/remove a single particle (linear combination)    
\end{verbatim}

In the future additional options will be added, such 
as two-nucleon spectroscopic factors, etc.

\section{Dot products of wave functions}

Option `{\tt (dotwfn)}' allows one to take the inner product (or dot product or overlap) between two wave functions, $\langle \psi_f | \psi_i \rangle$. This can also be done using {\tt BIGSTICK} 
with initial menu option `{\tt (v)}', but that requires 
both wave functions be in the same basis. In {\tt RHODIUM}, 
one can compute the dot product as long as both 
wave functions have the same $J_z$ (or $M$) and parity; 
they can be embedded in different Hilbert spaces, such as 
different truncations. One could, for example, find the 
overlap of a no-core shell-model $N_\mathrm{max}=6$ 
wave function with one with $N_\mathrm{max}=10$.

\section{Projection of a state into a basis 
with a different truncation}

Related to dot products, menu option `{\tt   (projct)}' 
allows one to project wave functions into a different 
Hilbert space, as long as the initial and final bases 
have the same $J_z$ (or $M$) and parity.  This is useful, 
for example, to project states from a smaller space into a 
much larger space. This can be useful for bootstrapping a block Lanczos calculation in {\tt BIGSTICK}, by using a block of wave functions computed in a small space as a pivot block (initial vectors) for a calculation in a larger space.

The output of this option is a {\tt BIGSTICK}-compatible file with extension {\tt .wfn}.

\section{Charge-conserving one-body density matrices}

If valence $N$ and $Z$ are the same, you get these 
options always (even if $M$ and parity $\pi$ are not the same):

\begin{verbatim}
  (dn1iso) 1-body density matrices, non-charge changing, good isospin 
  (dn1xpn) 1-body density matrices, non-charge changing, explicit proton-neutron     
\end{verbatim}

These options assume the sets of initial and final states are different. Indeed, you \textbf{cannot} 
use the same {\tt .wfn} files when choosing these options; if you want the same data, you must make a copy with a different name, but the same extension.
(Otherwise one can simply use {\tt BIGSTICK} to 
generate the density  matrices.)
The output will be a {\tt .dres} file.

You will be given an option to combine the two sets
of wave functions into a single list:
\begin{verbatim}
Do you want to combine initial/final levels into a single list? (y/n)    
\end{verbatim}

If you choose `{\tt n}', then 
the output will be a {\tt .dres} file, which lists both the initial and final sets of states, e.g.,
\begin{verbatim}
# Wavefunctions -- initial    4   4
  State       E          Ex           J        T     par 
    1      -87.10445     0.00000    -0.000   0.000   1
    2      -85.60215     1.50230     2.000   0.000   1
    3      -82.98830     4.11615     2.000   0.000   1
    4      -82.73202     4.37243     4.000   0.000   1
    5      -82.03408     5.07037     3.000   0.000   1
 # 
# Wavefunctions -- final      4   4
    1      -92.77905     0.00000    -0.000   0.000   1
    2      -91.11964     1.65940     2.000   0.000   1
    3      -88.47786     4.30119     2.000   0.000   1
    4      -87.97810     4.80094     4.000   0.000   1
    5      -87.43483     5.34422     3.000   0.000   1
    6      -86.54257     6.23647     4.000   0.000   1
    7      -84.79132     7.98772    -0.000   0.000   1
    8      -84.54140     8.23765     2.000   0.000   1
\end{verbatim}
This is slightly different from the standard 
{\tt BIGSTICK} output {\tt .dres} file, which provides a single lists of levels. If you choose `{\tt y}', that is, a single combined list, then the output {\tt .dres} file will generally be compatible with various postprocessing codes.

The output {\tt .dres} file also includes a convenient list of single-particle orbitals.
\begin{verbatim}
  Single particle state quantum numbers
ORBIT      N     L   2 x J  
     1     0     2     3
     2     0     2     5
     3     1     0     1    
\end{verbatim}
The output densities are in the same format 
as from {\tt BIGSTICK}, either in isospin format
\begin{verbatim}
 Initial state #    1 E =  -87.10445 2xJ, 2xT =    0   0
 Final state   #    1 E =  -92.77905 2xJ, 2xT =    0   0
 Jt =   0, Tt = 0        1 
    1    1   0.3876706   0.0000000
    2    2   1.7379967   0.0000000
    3    3   0.4368686   0.0000000    
\end{verbatim}
or in explicit proton 
neutron format,
\begin{verbatim}
 Initial state #    1 E =  -87.10445 2xJ, 2xT =    0   0
 Final state   #    1 E =  -92.77905 2xJ, 2xT =    0   0
 Jt =   0, proton      neutron 
    1    1   0.2741245   0.2741245
    2    2   1.2289493   1.2289492
    3    3   0.3089128   0.3089128    
\end{verbatim}
In both these cases, the initial wave vectors are drawn exclusively from the list of initial levels, and 
final from final.

\bigskip

Here is a sample script for obtaining cross-parity (i.e., between positive and negative parity levels). Note that one needs basis files for both parities, as well as wave function files for both parities.
\begin{verbatim}
n24pos.m0   ! INITIAL BASIS FILENAME
n24neg.m0   ! FINAL BASIS FILENAME
dn1xpn      ! MENU CHOICE: proton-neutron 1-body densities
y           ! combine positive, negative levels into one list
n24pos2neg  ! OUTPUT FILE NAME
n24pos      ! INITIAL WFN FILENAME
1 500       ! selected initial levels
n24neg      !  FINAL WFN FILENAME
1 500       ! selected intial levels
\end{verbatim}
Note time-reversed densities, that is, both positive-to-negative and negative-to-positive, are automatically generated.

\subsection{Generating `missing' density matrix elements}

In computing reduced density matrices, one must divide by a 
Clebsch-Gordan coefficient or an equivalent Wigner 3-$j$ symbol. 
However in some cases these vanish due to symmetries, as 
explained above. This means one is missing one-body density matrix elements $\langle J_f || [ \hat{c}^\dagger \otimes \tilde{c} ]_{K} || J_i \rangle $ when 
$(-1)^{J_i+J_f + K}= -1.$  (Note that here we use $K$ for the 
angular momentum rank of the operator to be consistent 
with the definition of reduced density matrices in Eq.~(\ref{reduced2}); however in {\tt .dres} file this is written 
as {\tt Jt}. 

{\tt RHODIUM} allows you to retrieve these `missing' matrix elements without re-running {\tt BIGSTICK}. By applying the angular momentum raising operator, $\hat{J}_+$, {\tt RHODIUM} can break the troublesome symmetry.

Recall this only happens if $M=0$ (and hence an even number of particles). If the initial basis and the final basis differ by exactly one unit of $M$, then you will see three additional options:
\begin{verbatim}
(jraise) Apply J+ raising operator to wave functions and write out 
(dn1jrm) Compute missing density matrix elements by applying J+ raising
(dn1mis) Compute missing density matrix elements     
\end{verbatim}

When seeking missing density matrix elements, there are two cases: first, when the list of initial and final states are the same, and second, when they are different.  In practice,
these two scenarios happen if the initial and final states have the same parity, or the opposite parities, respectively.

If the initial and final sets of states (levels) are the same, i.e., are drawn from the same parity, then use 
option `{\tt (dn1jrm)}.'  In this case, the original 
levels have been generated with $M=0$.  You will need to 
create basis files with extension {\tt .bas} with both 
$M=0$ and $M=1$.  With option `{\tt (dn1jrm)}', you will 
enter the filename for the wavefunctions computed with $M=0$;
{\tt RHODIUM} will them automatically apply the angular momentum raising operator $\hat{J}_+$, converting internally 
from $M=0$ to $M=1$ (levels with $J=0$ which, after application of $\hat{J}_\pm$, have zero norm and will be excluded; they don't contribute to missing matrix elements at any rate), and then compute a {\tt .dres} file with only the `missing' matrix elements, namely those that 
satisfy $(-1)^{J_i+J_f + K}= -1.$

\bigskip

Here is a sample script for obtaining `missing' density matrix elements when working from the same set of states (i.e., same parity):
\begin{verbatim}
n24pos.m0   ! INITIAL (INPUT) BASIS FILENAME
n24pos.m1   ! FINAL (TEMPORARY) BASIS FILENAME
dn1jrm      ! MENU OPTION
n24pos      ! INPUT FILENAME this is M = 0
1 500       ! selected levels
n24posmis   ! OUTPUT FILENAME
\end{verbatim}
Note that one needs the basis files for $M=0,1$, but one only 
reads in the $M=0$ wave function file. The $M=1$ basis file is 
needed as a reference, but the $M=1$ wave functions are automatically 
generated (and written temporarily to the file {\tt TEMP.wfn}) and then discarded.

\bigskip

If the initial and final sets of states (levels) are different, for example if they are drawn from different parities, then there is a two-step process: you must run {\tt RHODIUM} twice. This is because there are \textit{three} 
bases involved.  

Let's assume you start with two {\tt .wfn} files, one with basis $+, M=0$ and the other with basis $-, M=0$. 
In order to compute the missing density matrix elements, 
we have to put one of those sets of wave functions into 
a basis with $M=1$ (one could select instead $M=-1$; the choice does not matter), by applying the angular momentum 
raising operator $\hat{J}_+$, as 
\begin{equation}
    \hat{J_+} | M \rangle \propto | M+1 \rangle.
\end{equation}
Let's assume we choose to raise the - parity state; the choice does not matter. 

To being the process, one has to generate three {\tt .bas} files: $+,M=0$; $-,M=0$; and $-,M=1$. In the first step, 
in {\tt RHODIUM} take $-,M=0$ as the initial basis and $-,M=1$ as the final basis, and select option `{\tt (jraise)}'.
This option will request the input {\tt .wfn} file 
of wave vectors in the $-,M=0$ basis and then write out a new file those same wave vectors in the $-,M=1$ basis. The states will be normalized. Any state with $J=0$ has zero norm after 
raising (or lowering) and is ignored. 

In the second step, use the option `{\tt (dn1mis)}' to 
compute the missing density matrix elements between the 
states in the $+,M=0$ basis and the newly constructed states
in the $-,M=1$ basis; the order of initial and final does not matter. When asked
\begin{verbatim}
Do you want to combine initial/final levels into a single list? (y/n)
\end{verbatim}
answer `y`. This will produce a {\tt .dres} file in the standard format, but only containing the troublesome missing density matrix elements.

\bigskip

Here are sample scripts. For the first step, raising the negative 
parity states from $M=0$ to $M=1$. (Note: at the time of this writing, 
unfixed bugs require that the negative parity states get raised.)
\begin{verbatim}
negbasis.M0   
negbasis.M1
jraise         ! MENU CHOICE
inputwfn       ! must be negative parity, M = 0
1, 500
outputwfn      ! will be negative parity, M = 1
\end{verbatim}

The second step is to find the `missing' matrix elements, generated between positive parity, $M=0$ states and negative parity, $M=1$ states constructed in the previous step:
\begin{verbatim}
n24pos.m0
n24neg.m1
dn1mis           ! MENU CHOICE
y                ! COMBINE BOTH SETS OF LEVELS INTO ONE LIST
n24neg2posmis    ! OUTPUT FILENAME
n24pos           ! M = 0 WFN FILE
1 500           
n24negm1         ! M = 1 WFN FILE (from previous step)
1 500
\end{verbatim}
Again, time-reversed densities are automatically constructed. 
\textbf{As stated before, it's important to remember that we must have $J_t \geq |M| = |M_i-M_f|$}, 
where $J_t$ is the angular momentum rank of the transtion matrix elements. {\tt RHODIUM} will automatically skip such density matrix elements.

\bigskip

While these steps can seem complicated, they ultimate 
save compute time by reducing the costly Lanczos runs in {\tt BIGSTICK} to generate the wave functions. These procedures also eliminate ambiguities of phase and convergence, the latter of which can happen when a large number of states are generated.

\section{Applying a (non-charge-changing)
one-body operator}

This is option is very similar to the option `{\tt (o)}' in 
{\tt BIGSTICK}: the code reads in a wave function {\tt .wfn} file, and a {\tt .opme} file of reduced matrix elements specifying 
a one-body operator $\hat{O}$, then applies $\hat{O}$ 
to each of the wave vectors, and finally write  all the so modified wave functions  to a new 
{\tt .wfn} file.

The difference here is that the final states can have different basis quantum numbers from the initial states. 
For example, an E1 operator could be applied to + parity states to generate - parity states. As another example, 
a raising operator could take states from $M=0$ to $M=1$ 
(though {\tt RHODIUM} already has an option, {\tt (jraise)},
which does this automatically without having to read 
in the raising operator matrix elements). 

\section{Charge-changing one-body density matrices}

If the total number of nucleons is conserved, but 
$N$ and $Z$ both change by 1, {\tt RHODIUM} provides 
these options:
\begin{verbatim}
  (dn1bcc) Charge-changing 1-body densities (pn)
  (d1iscc) Charge-changing 1-body densities (iso)    
\end{verbatim}
These allow  you to compute charge-changing one-body density matrix elements, for example for beta decay, without assuming isospin is a good quantum number. In this case, the first option is preferred.

Note in this case, especially using the {\tt pn} option, to compute beta-decay matrix elements one 
cannot use the usual code {\tt gtstrength} but rather 
the {\tt pntrans} code.

\section{Spectroscopic amplitudes}

Spectroscopic amplitudes are reduced matrix elements of one-nucleon creation/annihilation 
operators, e.g.,
\begin{equation}
    {\cal A}_j^- = - \frac{ \langle A-1 || \hat{a}_j || A \rangle}{ \sqrt{2J_{A-1} +1 }},
\end{equation}
where $| A \rangle$ is the initial wave function with $A$ nucleons, $| A-1 \rangle$ is the 
final wave function with $A-1$ nucleons (and angular momentum $J_{A-1}$, and $\hat{a}_j$ is
the nucleon annihilation operator for the $j$th orbital.  Similarly,
\begin{equation}
    {\cal A}_j^+ = - \frac{ \langle A+1 || \hat{a}^\dagger_j || A \rangle}{ \sqrt{2J_{A+1} +1 }}.
\end{equation}
The reduced matrix elements are computed according to the conventions of Edmonds (\cite{edmonds1996angular}). 
In the literature, one more commonly uses the 
\textit{spectroscopic factor},
\begin{equation}
    S_j = ({\cal A}_j^+)^2.
\end{equation}

To compute the spectroscopic amplitude in {\tt RHODIUM}, you need to first create the following files using {\tt BIGSTICK}:
\begin{itemize}
    \item The {\tt .wfn} wave function file for the initial wave function (typically with option {\tt `(n)'}, though 
    other options can also create it);

    \item The {\tt .bas} basis file for the initial model space  using the option {\tt `(b)'};

    \item And the {\tt .wfn} and {\tt .bas} files for the final wave function. These must have one more or one less nucleon than the initial wave function.
    
\end{itemize}

As an example, let's suppose we've created files for $^{47,48}$Cr in the $pf$ valence space using 
the GX1A interaction. Starting {\tt RHODIUM}, we enter the names of the basis files, 
{\tt cr47.bas} and {\tt cr48.bas}, leaving off the mandatory extension:
\begin{verbatim}
      Enter name of INITIAL .bas file 
cr47
  Basis file successfully opened 
  about to read in basis
 dimbasischeck=               483887
.....
  DONE READING IN 
  Enter name of FINAL .bas file 
cr48
  Basis file successfully opened 
 dimbasischeck=              1963461
....
 Choose one of the following menu options 
   
  Enter the six-character code for your choice::
  (spamp1) 1-particle spectroscopic amplitude (change in number of particles)
  (appsp1) Add/remove a single particle (linear combination)
\end{verbatim}
Because {\tt RHODIUM} detected these two bases differ by exactly one nucleon, it offered
these two menu choices. In the next section we will consider the option {\tt (appsp1)}. Here
we consider the first option, {\tt (spamp1)}.

\begin{verbatim}
  You are calculating spectroscopic amplitudes with changes in the number of particles
  Enter output name (enter "none" if none)
Cr47toCr48
 ....    
  finished setting up local vectors...
  Initial state wave function file
  Enter input name of .wfn file 
cr47
 dimbasischeck=               483887
  .bas and .wfn headers agree 
  There are           10  initial wave functions 
  Enter start, stop to compute spectroscopic factors 
  (Enter 0,0 to take all )
0 0
    1      -83.81071     0.00000     2.500   0.500
    2      -83.78111     0.02959     1.500   0.500
    3      -83.69141     0.11930     3.500   0.500
    4      -82.56263     1.24808     5.500   0.500
...
  Final state wave function file
  Important: cannot be the same file as initial 
  Enter input name of .wfn file 
cr48
 dimbasischeck=              1963461
  .bas and .wfn headers agree 
  There are            5  final wave functions 
  Enter start, stop to compute spectroscopic factors 
  (Enter 0,0 to take all )
0 0
    1      -99.57887     0.00000     0.000   0.500
    2      -98.79021     0.78866     2.000   0.500
...
  
 All done with spectroscopic factors
\end{verbatim}
The output is written to a {\tt .spres} file, here {\tt Cr47toCr48.spres}.
That file has information on the initial and final wave functions:
\begin{verbatim}
  RHODIUM version 0.8.9 May 2024 
# Wavefunctions -- initial    4   3
    1      -83.81071     0.00000     2.500   0.500
    2      -83.78111     0.02959     1.500   0.500
....
 # 
# Wavefunctions -- final      4   4
    1      -99.57887     0.00000     0.000   0.500
    2      -98.79021     0.78866     2.000   0.500
...
\end{verbatim}
and on the single particle orbitals
\begin{verbatim}
  Single particle state quantum numbers
ORBIT      N     L   2 x J  
     1     0     3     7
     2     1     1     3
     3     0     3     5
     4     1     1     1    
\end{verbatim}
and finally on the single particle amplitudes themselves:
\begin{verbatim}
 Initial state #    1 E =  -83.81071 2xJ, 2xT =    5   0
 Final state   #    1 E =  -99.57887 2xJ, 2xT =    0   0
 orbit  amp 
    3  -0.353712
  
 Initial state #    1 E =  -83.81071 2xJ, 2xT =    5   0
 Final state   #    2 E =  -98.79021 2xJ, 2xT =    4   0
 orbit  amp 
    1   0.697354
    2   0.227033
    3  -0.083912
    4  -0.061296
....    
\end{verbatim}
The code will automatically output the amplitude for proton or neutron; it's up to the user to remember which species it is.

At the end of the {\tt .spres} file, the code also writes out sum rules (to be finished).

\subsection{Sum rules}

A check on the spectroscopic amplitudes are sum rules.

We define the spectroscopic amplitude for adding a particle (there is actually a different definition) as 
\begin{equation}
A^{fi,+}_j = \frac{  \langle J_f || \hat{a}^\dagger_j || J_i \rangle }{ [J_f] }.
\end{equation}
One can show (\cite{rowe2010fundamentals}) that 
\begin{equation}
A^{fi,+}_j = - \langle J_f \, M_f | \left [ \hat{a}^\dagger_j \otimes | J_i \rangle \right ]_{J_f \, M_f },
\end{equation}
independent of $M_f$.  With a bit of work, one can show that 
\begin{equation}
\sum_i \left (A^{fi,+}_j \right )^2 =  \langle f | \hat{n}_j | f \rangle,
\end{equation}
where $\hat{n}_j$ is the number operator for the orbital $j$ (it should be apparent I am suppressing all other quantum numbers on the orbital).  We can sum the other way:
by taking particular care with
the sum over Clebsch-Gordan coefficients, one can show that 
\begin{equation}
\sum_f \left (A^{fi,+}_j \right )^2 \frac{2J_f +1}{2J_i+1}  = 2j+1 - \langle i | \hat{n}_j | i \rangle.
\end{equation}
As $2j+1$ is the maximum number of particles in this orbital, this sum rule counts how many holes there are in the orbital $j$. 

These are sum rules for \textit{adding} a particle to an initial state. What about removing a particle?  By using the definition of reduced matrix elements, one can show
\begin{equation}
\langle J_f || \hat{a}^\dagger_j || J_i \rangle = (-1)^{J_f + j - J_i } \langle J_i || \tilde{a}_j || J_f \rangle .
\end{equation}
If we define
\begin{equation}
A^{fi,-}_j = \frac{  \langle J_f || \tilde{a}_j || J_i \rangle }{ [J_f] },
\end{equation}
then 
\begin{equation}
A^{fi,-}_j  = (-1)^{J_i + j - J_f } \frac{ [J_i]}{[J_f]} A^{if,+}_j.
\end{equation}
With this we then get the sum rules 
\begin{equation}
\sum_i \left (A^{fi,-}_j \right )^2 = 2j+1 - \langle f | \hat{n}_j | f \rangle
\end{equation}
and
\begin{equation}
\sum_f \left (A^{fi,-}_j \right )^2 \frac{2J_f +1}{2J_i+1}  =\langle i | \hat{n}_j | i \rangle.
\end{equation}

\section{Adding or removing a nucleon}

Rather than choose the {\tt `(spamp1)'} option to compute the spectroscopic amplitude, one can choose the option 
\begin{verbatim}    
  (appsp1) Add/remove a single particle (linear combination)
\end{verbatim}
which constructs new wave function(s) by adding or removing a single nucleon, or a  linear combination of nucleons--for example, a plane wave, and writing that to a {\tt .wfn} file.

To do that you need a file with the single-particle amplitudes, with extension {\tt .spa}. The {\tt RHODIUM} distribution comes with an example, {\tt SAMPLE.spa}:
\begin{verbatim}
#SAMPLE SPA FILE
3   ! number of orbitals
1   0  2  1.5  ! index   n  l   j
2   0  2  2.5
3   1  0  0.5
2       ! species = 1 protons 2 neutrons
1  -0.1  ! orbital amplitude
2   1.3
3   0.05
\end{verbatim}
The file includes:
\begin{enumerate}

    \item An optional header, started by \# or !

    \item The number of orbitals (assumed to be the same for protons and neutrons)

    \item A list of the orbitals, including (a) the index of the orbital (b) the integer radial nodal 
    number (c) the integer orbital angular momentum $l$ (d) the real total angular momentum $j$

\item The species, = 1 for protons, = 2 for neutrons (this must be specified, as I assume that protons could be affected by Coulomb)

\item Finally, a list of the amplitude $c_i$ for each orbital index $i$

\end{enumerate}
This will create a wave function, e.g., for creating a nucleon:
\begin{equation}
    | \Psi_{A+1} \rangle = \left (   \sum_i  c_i \hat{a}^\dagger_i \right ) | \Psi_A \rangle.
\end{equation}
{\tt RHODIUM} automatically takes care of $m$-values, and uses the input file the same for creating or annihilating a nucleon.

After reading in the {\tt .spa} amplitude file, you will be asked if you want to normalize 
the wave function or not:
\begin{verbatim}
  Do you want to normalize final wfns? (y/n)    
\end{verbatim}
The next step is to read in the initial {\tt .wfn} file previously generated (either by {\tt BIGSTICK} or by a previous application of {\tt RHODIUM}).
\begin{verbatim}
  There are         N  initial wave functions 
  Enter start, stop to apply one particle creation/removal
  (Enter 0,0 to take all )    
\end{verbatim}
Finally you will be asked the name of the output file (with extention {\tt .wfn}, but do not 
include the extension)
\begin{verbatim}
  Enter name of output .wfn file 
test    
\end{verbatim}

\bigskip

The option {\tt appsp1} produces only a single final state in the output {\tt .wfn} file.  This can be inconvenient. 
For example, in generating an optical potential using Green's functions, one may want a nucleon added/removed for each of $0d_{5/2}, 0d_{3/2}, 1s_{1/2}$. Using only option {\tt appsp1}, 
to add/remove a nucleon systematically from each available orbital, one would have to run the code multiple times, once for each orbital. In the $sd$-shell, this would require three files, one for each 
of $0d_{5/2}, 0d_{3/2}, 1s_{1/2}$.

%%%%% REMOVED SECTION ON OPTION allsp1

\chapter{Acknowledgements}

This material is based upon work supported by the U.S. Department of Energy, Office of Science, Office of Nuclear Physics, 
under Award Number  DE-FG02-03ER41272.  I thank Mark Caprio for pointing out some inconsistencies and infelicities in the text.

\bibliographystyle{abbrvnat}
\bibliography{rhodium}

\end{document}